\documentclass{article}

\usepackage[T1]{fontenc}
\usepackage{graphicx}

\begin{document}

\title{
\textbf{First-Principles Calculation of Electronic Excitations in Solids with SPEX}
}

\author{
{\normalsize Arno Schindlmayr$^{1,}$\thanks{Corresponding author. E-mail: Arno.Schindlmayr@uni-paderborn.de} , Christoph Friedrich$^2$, Ersoy \c{S}a\c{s}{\i}o\u{g}lu$^2$, and}\\[-2pt]
{\normalsize Stefan Bl\"ugel$^2$}\\
{\small $^1$ Department Physik, Universit\"at Paderborn, 33095 Paderborn, Germany}\\[-4pt]
{\small $^2$ Institut f\"ur Festk\"orperforschung and Institute for Advanced Simulation,}\\[-4pt]
{\small Forschungszentrum J\"ulich, 52425 J\"ulich, Germany}
}

\date{}

\maketitle

%Keywords: Many-Body Perturbation Theory / Self-Energy / Quasiparticle Band Structure / Dielectric Function / Spin Waves

\begin{abstract}
\noindent We describe the software package SPEX, which allows first-principles calculations of quasiparticle and collective electronic excitations in solids using techniques from many-body perturbation theory. The implementation is based on the full-potential linearized augmented-plane-wave (FLAPW) method, which treats core and valence electrons on an equal footing and can be applied to a wide range of materials, including transition metals and rare earths. After a discussion of essential features that contribute to the high numerical efficiency of the code, we present illustrative results for quasiparticle band structures calculated within the $GW$ approximation for the electronic self-energy, electron-energy-loss spectra with inter- and intraband transitions as well as local-field effects, and spin-wave spectra of itinerant ferromagnets. In all cases the inclusion of many-body correlation terms leads to very good quantitative agreement with experimental spectroscopies.
\end{abstract}

\section{Introduction}

First-principles computations, which do not rely on any empirical input parameters, have become an important tool in materials science. Ideally, such calculations use only the fundamental laws of nature together with the specified elemental composition in order to predict the structural and electronic properties of a material. As chemical bonding and the response to external fields are determined by the microscopic dynamics of the ions and electrons inside the solid, the relevant laws in this case are those of quantum mechanics. The complete Hamiltonian can, in fact, be readily written down, because the interaction potentials between all constituent particles, given by Coulomb's law, are known exactly. However, a direct solution of the Schr\"odinger equation or its relativistic counterpart, the Dirac equation, is not feasible for extended solids, because the number of electrons $N$ is of the order of $10^{23}/\mathrm{cm}^{3}$ and hence too large for a numerical treatment of the correlated many-electron wave function $\Psi(\mathbf{r}_1,\ldots,\mathbf{r}_N)$. Instead, practical applications rely on alternative approaches that are formally equivalent to the Schr\"odinger equation but do not employ the many-electron wave function as the basic variable. A prominent example is density-functional theory (DFT), which is based on the ground-state electron density $n(\mathbf{r})$, a real quantity that depends only on a single spatial position vector. In spite of this vast simplification, the Hohenberg-Kohn theorem \cite{Hohenberg1964} asserts that knowledge of the ground-state density alone is sufficient, at least in principle, to determine all observables of a stationary system in equilibrium. The density itself can be calculated from the Kohn-Sham scheme \cite{Kohn1965}, which introduces an auxiliary system of non-interacting electrons with the same ground-state density as the real interacting system and requires the solution of a single-particle Schr\"odinger-like equation, the Kohn-Sham equation, with a self-consistent local potential.

Despite the enormous scope of the Hohenberg-Kohn theorem, practical applications are limited because for most observables no explicit formulas for the actual dependence on $n(\mathbf{r})$ are known. A notable exception, besides the density itself, is the ground-state total energy, where the dominant contributions are given exactly and the remaining exchange-correlation functional may be replaced by the local-density approximation (LDA) \cite{Kohn1965} or generalized gradient approximations like PBE \cite{Perdew1996a}. By comparing the total energies for different atomic configurations it is thus possible to predict crystal structures and related quantities like the elastic moduli without empirical parameters. The huge success of density-functional theory stems from the good agreement with crystallographic measurements for a wide range of different materials. In contrast, electronic excitation spectra often show substantial deviations from experimental spectroscopies. The most famous example is the severe underestimation of the band gaps of semiconductors. As a large part of the error in this case is systematic and arises from the pervasive but incorrect identification of the Kohn-Sham eigenvalues with the true quasiparticle energies, better approximations for the exchange-correlation functional will not solve the problem. Therefore, quantitative methods for electronic excitations in solids are now chiefly sought outside density-functional theory. Perhaps the simplest extension are hybrid functionals like PBEh \cite{Perdew1996b} or HSE \cite{Heyd2003}, which replace a fraction of the local exchange potential by non-local Hartree-Fock exchange. As Hartree-Fock in turn tends to overestimate band gaps, a linear combination with suitably chosen coefficients can improve the agreement with experiments. Unfortunately, the optimal relative weights depend on the material: While an admixture of 25\% Hartree-Fock yields good results for many typical semiconductors, a higher fraction is required for large-gap insulators, where screening is much weaker \cite{Paier2006}. For metals any inclusion of Hartree-Fock exchange even increases the already too large valence-band widths further, so that hybrid functionals are in fact in worse agreement with experimental band structures than pure LDA results \cite{Paier2006}.

A better-founded scheme without adjustable parameters is many-body perturbation theory \cite{Mahan1990}, which is based on the Green function $G(\mathbf{r},\mathbf{r}';\omega)$. As it contains no inbuilt systematic errors, the accuracy is only limited by functional approximations like the $GW$ approximation for the electronic self-energy \cite{Hedin1965}, which can be improved if necessary. As a consequence, this approach yields excitation energies in significantly better agreement with experiments than density-functional theory. The main drawback is the high computational cost, which is related to the more complicated mathematical form of the non-local and frequency-dependent Green function in comparison to the density. For this reason, practical applications have so far been limited to moderately complex systems with up to about one hundred atoms per unit cell. Although this suffices to study, for example, carbon nanotubes \cite{Spataru2004} or point defects at semiconductor surfaces \cite{Hedstrom2002}, simulations with tens of thousands of atoms are now possible using density-functional theory with linear-scaling algorithms \cite{Goedecker1999}. Furthermore, many actual $GW$ calculations contain a number of additional simplifications.

Besides individual quasiparticle properties, collective excitations of the electron system in solids constitute another major challenge for electronic-structure calculations. Prominent examples are plasmons and excitons, which are associated with resonances in the dielectric function and can be probed by electron-energy-loss or optical spectroscopies, but also spin-wave modes in magnetic materials. As a single-particle picture cannot describe such collective excitations, many-body perturbation theory has become the method of choice for quantitative simulations \cite{Onida2002}.

Here we describe a new software package, SPEX \cite{SPEX}, which contains an implementation of many-body perturbation theory and can be used to simulate various spectroscopic techniques. The code is designed to avoid many shortcomings of previous implementations and to keep additional approximations at a minimum. Most importantly, it is not based on the prevalent pseudopotential concept but uses the full-potential linearized augmented-plane-wave (FLAPW) method \cite{Singh1994}, which opens the door to a wider range of materials, including elements with localized \textit{d} or \textit{f} orbitals. In the following we first discuss some essential features of the code. Then we present selected results for several spectroscopies that probe electronic excitations, such as quasiparticle band structures, electron energy loss or spin waves, in order to illustrate the possible applications. Finally, we briefly summarize and comment on planned future developments.

\section{The SPEX code}

Although many-body perturbation theory is a self-contained mathematical framework, actual applications are most efficient when combined with density-func\-tio\-nal theory. The Kohn-Sham eigenvalues, which are often in qualitative agreement with the true band structure, then serve as the starting point for a diagrammatic expansion in terms of the dynamically screened Coulomb interaction $W(\mathbf{r},\mathbf{r}';\omega)$. The screening arises from the formation of exchange-correlation holes around charged fermions and leads to the concept of quasiparticles, which incorporate the polarization of the local environment and constitute one class of elementary excitations of the electron system. As the screened interaction between the quasiparticles is much weaker than the bare Coulomb potential, a perturbative treatment is justified. A first-order expansion of the self-energy leads to the $GW$ approximation, which is just the linear term in $W(\mathbf{r},\mathbf{r}';\omega)$. In accordance with standard perturbation theory, its matrix elements must be evaluated with the original unperturbed Kohn-Sham orbitals.

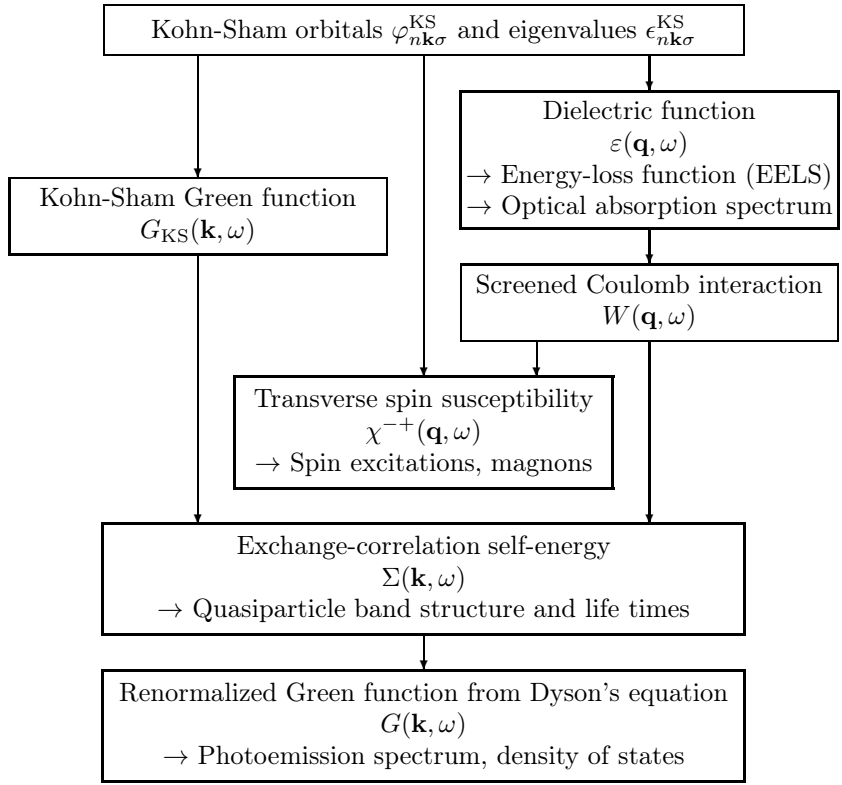
\begin{figure}
\setlength{\unitlength}{1cm}
\begin{picture}(11.0,10.3)
\put(1.25,9.65){\framebox(8.5,0.65){\parbox{9cm}{\centering
Kohn-Sham orbitals $\varphi^\mathrm{KS}_{n \mathbf{k} \sigma}$ and eigenvalues $\epsilon^\mathrm{KS}_{n \mathbf{k} \sigma}$
}}}
\put(2.5,9.65){\vector(0,-1){1.65}}
\put(0.0,7.0){\framebox(5.0,1.0){\parbox{5cm}{\centering
Kohn-Sham Green function\\
$G_\mathrm{KS}(\mathbf{k},\omega)$
}}}
\put(8.5,9.65){\vector(0,-1){0.5}}
\put(6.0,7.35){\thicklines \framebox(5.0,1.8){\parbox{5cm}{\centering
Dielectric function\\
$\varepsilon(\mathbf{q},\omega)$\\
$\rightarrow$ Energy-loss function (EELS)\\
$\rightarrow$ Optical absorption spectrum
}}}
\put(8.5,7.35){\vector(0,-1){0.5}}
\put(6.0,5.85){\framebox(5.0,1.0){\parbox{5cm}{\centering
Screened Coulomb interaction\\
$W(\mathbf{q},\omega)$
}}}
\put(5.5,9.65){\vector(0,-1){4.3}}
\put(7.0,5.85){\vector(0,-1){0.5}}
\put(3.0,3.9){\thicklines \framebox(5.0,1.45){\parbox{5cm}{\centering
Transverse spin susceptibility\\
$\chi^{-+}(\mathbf{q},\omega)$\\
$\rightarrow$ Spin excitations, magnons
}}}
\put(2.5,7.0){\vector(0,-1){3.6}}
\put(8.5,5.85){\vector(0,-1){2.45}}
\put(1.25,1.95){\thicklines \framebox(8.5,1.45){\parbox{9cm}{\centering
Exchange-correlation self-energy\\
$\Sigma(\mathbf{k},\omega)$\\
$\rightarrow$ Quasiparticle band structure and life times
}}}
\put(5.5,1.95){\vector(0,-1){0.5}}
\put(1.25,0.0){\thicklines \framebox(8.5,1.45){\parbox{9cm}{\centering
Renormalized Green function from Dyson's equation\\
$G(\mathbf{k},\omega)$\\
$\rightarrow$ Photoemission spectrum, density of states
}}}
\end{picture}
\caption{Flowchart for the perturbative calculation of different electronic excitations and spectroscopies as implemented in SPEX.}
\vspace{-0.04cm}
\label{Fig:flowchart}
\end{figure}

Following this philosophy, SPEX is designed as a separate module that computes electronic excitation spectra from a given set of Kohn-Sham orbitals and eigenvalues. It can be combined with any density-functional code whose output data is convertible to the FLAPW form. We use our own package FLEUR \cite{SPEX}, but other choices are equally possible. The flowchart in Fig.~\ref{Fig:flowchart} illustrates the course of the calculation.

A premier goal during the code development was to avoid additional simplifications wherever possible, so that the results depend exclusively on controllable convergence parameters and the choice of functional approximations like the $GW$ approximation for the self-energy or the random-phase approximation (RPA) for the dielectric function. Besides, emphasis was placed on high efficiency, especially in terms of CPU time, so that complex materials with large unit cells can be treated. Finally, the code should be versatile and easily adaptable to new physical problems, for instance in the emerging areas of nanomagnetism and spintronics. To satisfy these requirements SPEX contains a number of important features that are summarized below.

(i) Consistent employment of FLAPW, which treats core and valence electrons on the same footing. Unlike the atomic-sphere approximation used in early all-electron codes, the full-potential treatment is also suitable for surfaces or defects. The linearization error caused by expanding the muffin-tin functions around fixed energy parameters, which is negligible for states close to the Fermi level but becomes important for high unoccupied bands that contribute to the self-energy, can be eliminated by including higher energy derivatives as additional local orbitals \cite{Friedrich2006}.

(ii) The FLAPW basis set is optimized for the Kohn-Sham orbitals but does not span the complete Hilbert space. Representing products of wave functions, for example in the polarizability or matrix elements of the Coulomb potential, in this basis hence implies a loss of accuracy. Instead, we use a mixed product basis \cite{Kotani2002} that is constructed from products of the basis functions and allows exact projections. After linear dependencies are removed, this set may be truncated in a controlled fashion, if desired.

(iii) The dielectric function is constructed either in the random-phase or the time-dependent local-density approximation, without recourse to plasmon-pole models. The full frequency dependence as well as local-field effects are thus properly included. Besides, it gives access to the imaginary part of the self-energy, which leads to complex quasiparticle energies and describes the finite lifetime of the excited states due to scattering.

(iv) Frequency convolutions are normally evaluated by means of contour integrations in the complex plane, although the analytic continuation of functions calculated on the imaginary axis to real frequencies, which is faster but less well controlled, is also possible.

(v) Symmorphic and non-symmorphic spatial symmetries are exploited wherever possible. For systems with inversion symmetry we also use the fact that the Coulomb matrix and response functions on the imaginary frequency axis may be chosen real and thus processed in compact form.

(vi) Following a procedure developed in Ref.~\cite{Freysoldt2007} for plane waves, the singularity of the Coulomb matrix at $\mathbf{k} = \mathbf{0}$ in reciprocal space is treated exactly through an expansion that separates the divergent and regular parts \cite{Friedrich2009}. A subsequent diagonalization confines the singularity to a single divergent eigenvalue, which can be processed analytically. At the same time, the transformation from the mixed product basis to the eigenvectors of the Coulomb matrix allows a very efficient truncation by eliminating the least significant scattering channels with the smallest eigenvalues.

(vii) The spin degree of freedom is fully supported. We make no simplifying assumptions about the orbitals in the two spin channels and allow a completely spin-unrestricted treatment of magnetic materials.

(viii) The code is applicable both to insulators and to metals. Where intraband transitions require a special treatment, appropriate provisions are made. In particular, a Drude term is included in the dielectric function in this case. All calculations are done at zero temperature.

(ix) Relativistic corrections are treated at the scalar-relativistic level for valence states and the full Dirac equation for core states.

As a performance test we conducted $GW$ band-structure calculations for diamond supercells of various sizes, using a $\mathbf{k}$-point sampling equivalent to 4$\times$4$\times$4 mesh points in the full Brillouin zone corresponding to the elementary diatomic unit cell and with optimized parameters that otherwise guarantee a convergence of the quasiparticle shifts to within 0.01\,eV \cite{Friedrich2010}. Our results demonstrate that simulations even with 128 atoms per cell are perfectly feasible on a standard single-processor work station, and a complete quasiparticle band structure requires less than one and a half days of CPU time in this case. We find a scaling behavior that lies between quadratic and cubic with the number of atoms in this size range.

\section{Results}

In this section we present illustrative results for different spectroscopies. All materials discussed here contain transition-metal elements and would be difficult to treat with conventional pseudopotential codes, because the localized \textit{d} orbitals require a large number of plane waves.

\subsection{Quasiparticle band structures}

\begin{figure}[t!]
\centering \includegraphics*[scale=0.35]{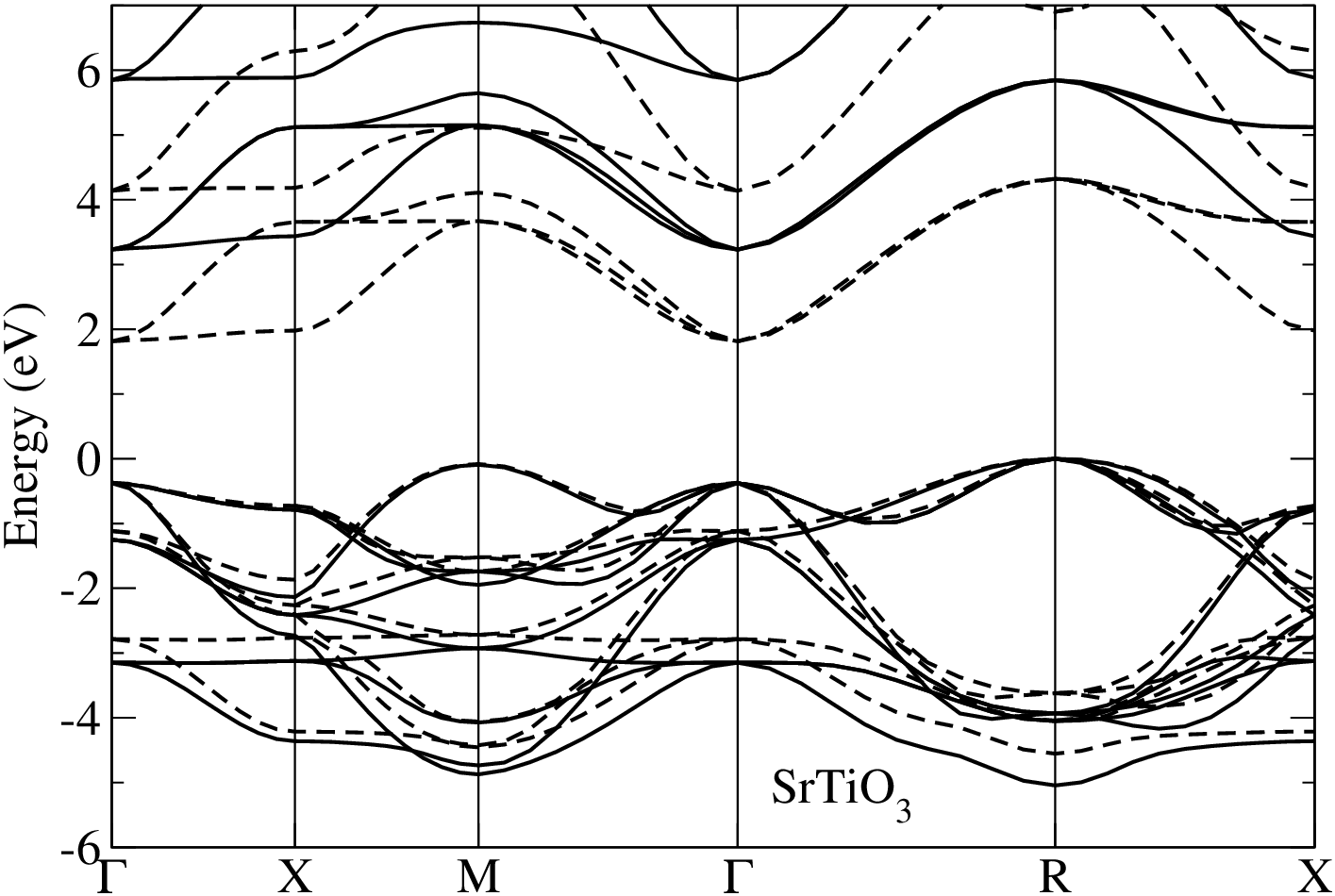}
\caption{Quasiparticle band structure of cubic SrTiO$_3$ calculated in DFT-PBE (dashed lines) and the $GW$ approximation (solid lines).}
\label{Fig:STO-band}
\end{figure}

In the perturbative approach adopted here the quasiparticle energies are derived from the Kohn-Sham eigenvalues and the self-energy according to
\begin{equation}
\epsilon_{n \mathbf{k} \sigma} = \epsilon^\mathrm{KS}_{n \mathbf{k} \sigma} + \langle \varphi^\mathrm{KS}_{n \mathbf{k} \sigma} | \Sigma_\sigma(\epsilon_{n \mathbf{k} \sigma}) - V^\mathrm{xc}_\sigma | \varphi^\mathrm{KS}_{n \mathbf{k} \sigma} \rangle \;.
\end{equation}
The self-energy is evaluated in the $GW$ approximation, and the matrix elements of the local exchange-correlation potential $V^\mathrm{xc}_\sigma(\mathbf{r})$ are subtracted to avoid double counting. Although $GW$ calculations for real materials have been feasible since the 1980s \cite{Hybertsen1985,Godby1986}, the prevailing reliance on plane waves and pseudopotentials meant that applications were long restricted almost exclusively to \textit{sp}-bonded semiconductors and simple metals. No such restrictions apply to the FLAPW method. As an example, in Fig.\ \ref{Fig:STO-band} we show the band structure of strontium titanate (SrTiO$_3$), a high-$\kappa$ dielectric insulator from the perovskite family, in the cubic phase observed at room temperature. The lattice constant within DFT-PBE \cite{Perdew1996a} is 7.46\,Bohr. For the self-energy construction we use 64 mesh points in the full Brillouin zone, 550 unoccupied bands, and cutoff parameters $G_\mathrm{max} = 5\,\mathrm{Bohr}^{-1}$ and $L_\mathrm{max} = 4$ for the mixed product basis in the interstitial region and in the muffin-tin spheres, respectively. Second energy derivatives are included as additional local orbitals. Although DFT-PBE gives the correct qualitative picture, both the indirect band gap of 1.81\,eV between R and $\Gamma$ and the direct gap of 2.18\,eV at $\Gamma$ are significantly too low. In contrast, the $GW$ approximation yields 3.23\,eV and 3.61\,eV in very good agreement with the experimental values 3.25\,eV and 3.75\,eV \cite{vanBenthem2001}. The size of the band gap is quite important in this case, as SrTiO$_3$ has long been technically used as an optically transparent synthetic diamond simulant. The comparison with a previously published $GW$ value of 5.07\,eV for the indirect gap \cite{Cappellini2000}, obtained with a parametrized model dielectric function, underlines the need for an accurate evaluation of the self-energy correction.

\begin{figure}[b!]
\centering \includegraphics*[scale=0.35]{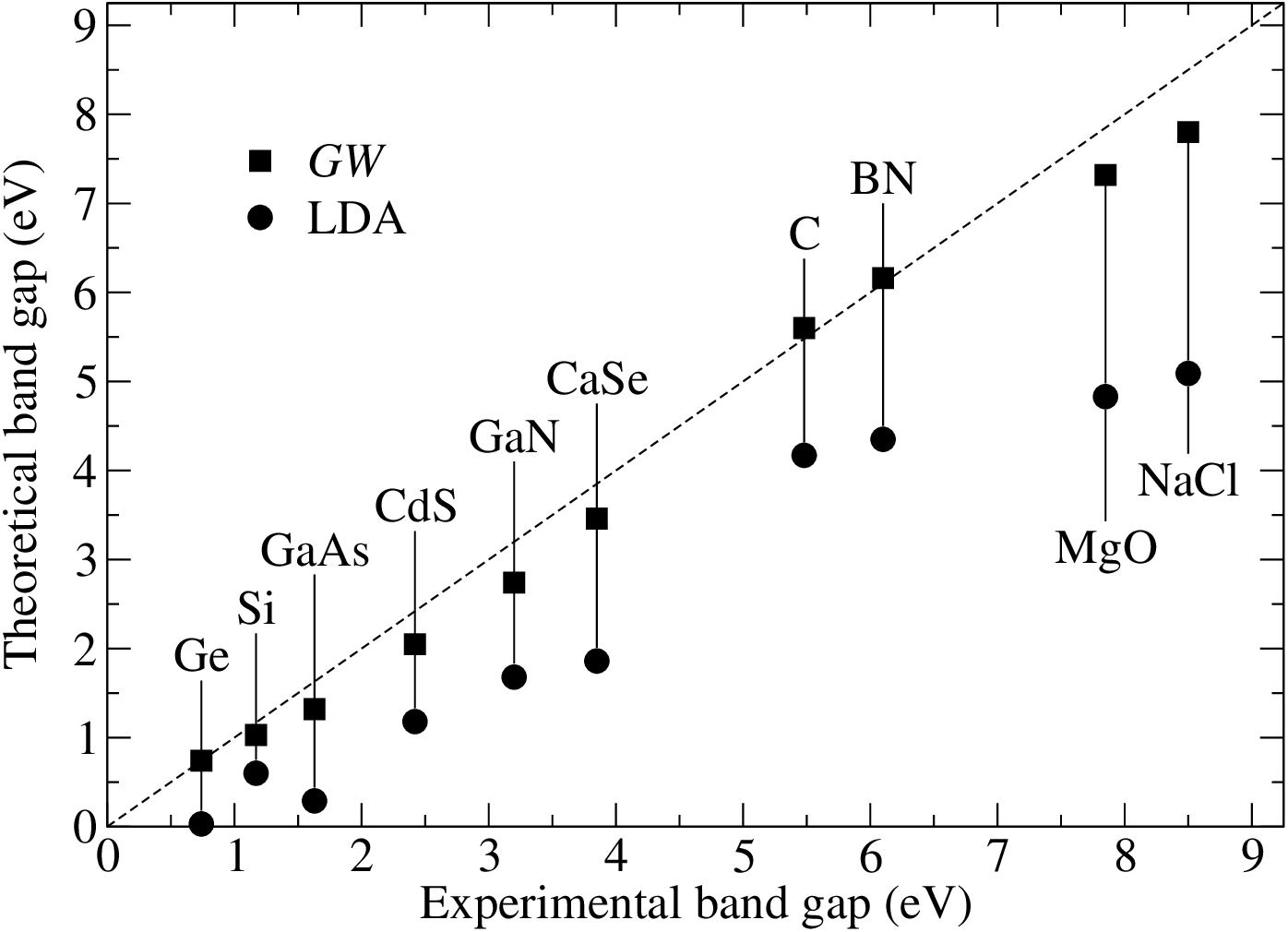}
\caption{Calculated band gaps of selected materials in the LDA (circles) and the $GW$ approximation (squares) compared to experimental values.}
\label{Fig:bandgaps}
\end{figure}

In Fig.~\ref{Fig:bandgaps} we display results for a wider range of materials from small-gap to large-gap insulators. In all cases the $GW$ approximation corrects virtually the entire error of the LDA\@. The data support our previous conclusion, based on an in-depth study of silicon \cite{Friedrich2006}, that carefully performed all-electron $GW$ calculations yield very good agreement with experiments. This had initially been in doubt after an early FLAPW calculation found unexpectedly large deviations both from experiments and from established pseudopotential values \cite{Ku2002}. However, this observed discrepancy is now understood to have arisen from incomplete convergence, especially with respect to the number of unoccupied conduction bands in the self-energy \cite{Friedrich2006,Tiago2004}. Our own results are meanwhile confirmed by other independent all-electron implementations \cite{Shishkin2006,vanSchilfgaarde2006}.

Incidentally, these studies also revealed that well converged all-electron $GW$ band gaps do not coincide with pseudopotential values after all. The remaining difference stems from the pseudization of the wave functions and the linearized core-valence interaction in the latter approach \cite{Gomez-Abal2008}.

\subsection{Electron-energy-loss spectroscopy}

\begin{figure}[b!]
\centering \includegraphics*[scale=0.35]{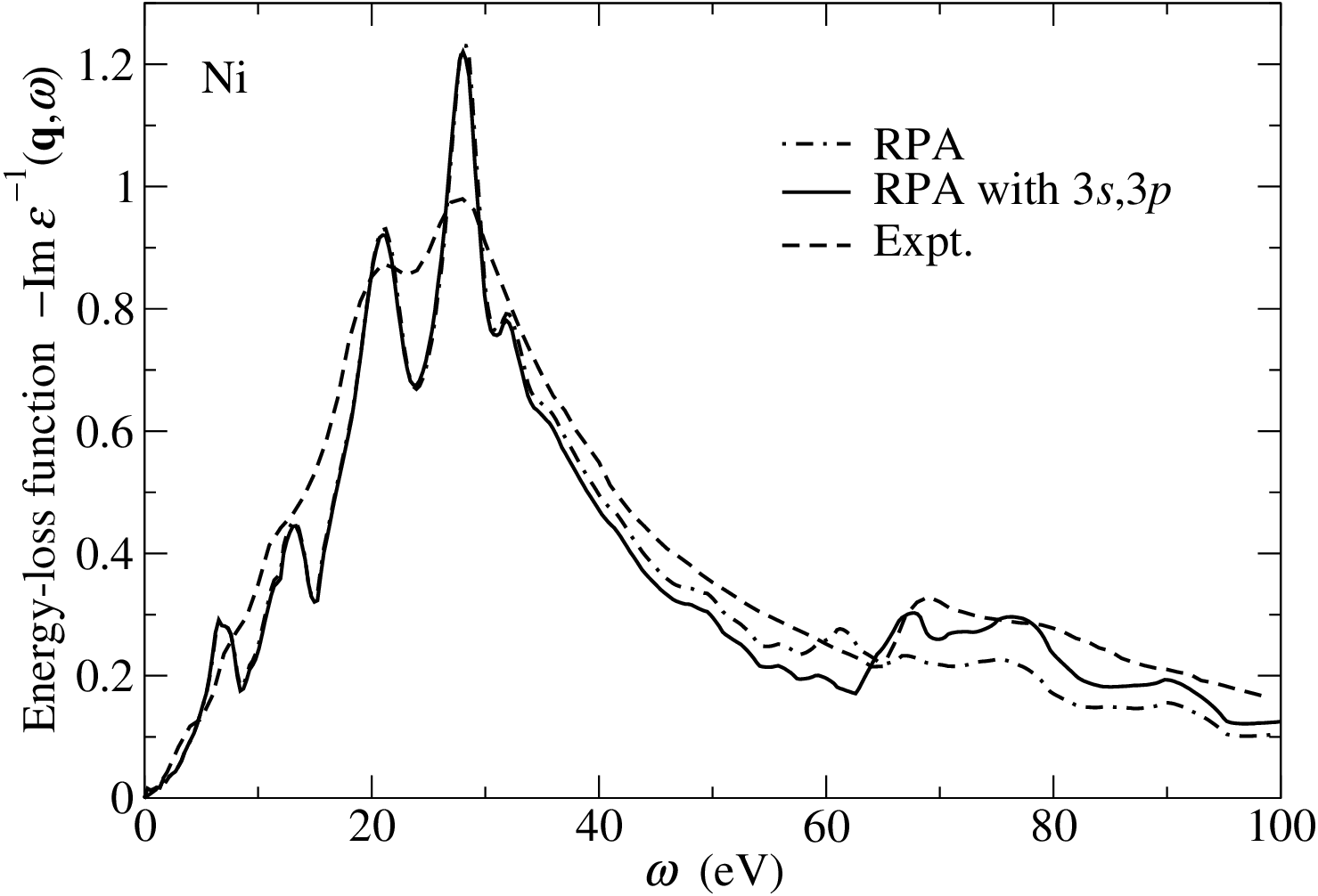}
\caption{Electron-energy-loss spectrum (EELS) of ferromagnetic Ni for $\mathbf{q} = (0.25,0,0) 2 \pi / a$ calculated in the RPA with (solid line) and without (dot-dashed line) transitions from the 3\textit{s} and 3\textit{p} core states compared to experimental data (dashed line) from Ref.~\cite{Feldkamp1979}.}
\label{Fig:EELS-Ni}
\end{figure}

While the single-particle Green function and self-energy describe the final states of photoemission experiments, where the particle number changes due to electron emission or injection, low-energy spectroscopies involving intraband or interband transitions are related to the dielectric function $\varepsilon(\mathbf{r},\mathbf{r}',\omega)$, which is linked to the density-density correlation function and characterizes the linear response to an external electric field. In particular, electron-energy-loss spectroscopy (EELS) measures the imaginary part of the inverse macroscopic dielectric function, the so-called energy-loss function $-\mathop{\mathrm{Im}} \varepsilon^{-1}(\mathbf{q},\omega)$. In Fig.\ \ref{Fig:EELS-Ni} we display the EELS spectrum of ferromagnetic nickel calculated in the RPA\@. The response function is constructed with 40$\times$40$\times$40 mesh points in the full Brillouin zone, $G_\mathrm{max} = 5\,\mathrm{Bohr}^{-1}$, $L_\mathrm{max} = 4$ and 118 unoccupied bands \cite{Friedrich2006}. Second and third energy derivatives are included as local orbitals. The resulting curve is in very good agreement with experimental measurements \cite{Feldkamp1979}, especially if transitions from the 3\textit{s} and 3\textit{p} core orbitals are taken into account; in this case the step in the energy-loss function around 64\,eV, which corresponds to the onset of transitions from these states, is also well reproduced. The RPA formally corresponds to a complete neglect of dynamic exchange-correlation effects in the linear density response function. The latter can be approximately included by the adiabatic local-density approximation \cite{Botti2007}, but in most cases the results change very little compared to the RPA\@.

\subsection{Spin-wave spectra}

Electric fields couple to the charge of the electrons and induce characteristic excitations, such as interband optical transitions or collective plasmon modes, which correspond to resonances in the dielectric function and can be measured with frequency-resolved spectroscopies. Likewise, magnetic fields couple to the spin of the electrons and give rise to excitations of the spin system. These also fall into two groups, spin-flip Stoner excitations of individual electrons and collective spin waves, and can be identified as resonances in the dynamic transverse spin susceptibility.

Most theoretical studies of spin waves are based on the Heisenberg model, although the assumption of localized spins makes its justification doubtful for ferromagnetic metals with itinerant electrons. In contrast, very few attempts at first-principles calculations were reported so far \cite{Savrasov1998,Karlsson2000}. The difficulty is twofold: As magnetic behavior originates in localized \textit{d} or \textit{f} orbitals of transition-metal and rare-earth elements, an all-electron scheme is mandatory. Furthermore, the treatment of spin waves requires dynamic exchange-correlation effects not contained in the RPA\@. We have solved the latter by explicitly including the screened Coulomb interaction between electrons and holes in the two different spin channels \cite{Aryasetiawan1999}. The resulting two-particle problem is analogous to the Bethe-Salpeter equation used for excitons in semiconductors. As the dominant part of the dynamic correlation in ferromagnets is caused by multiple scattering of electron-hole pairs at the same atomic site, we transform the screened interaction to a basis of maximally localized Wannier functions \cite{Freimuth2008}, which allows a very efficient truncation. Here we take only matrix elements with four Wannier functions that are all localized at the same site into account, but systematic extensions are of course possible to ensure full convergence \cite{Sasioglu2010}.

\begin{figure}[t!]
\centering \includegraphics*[scale=0.35]{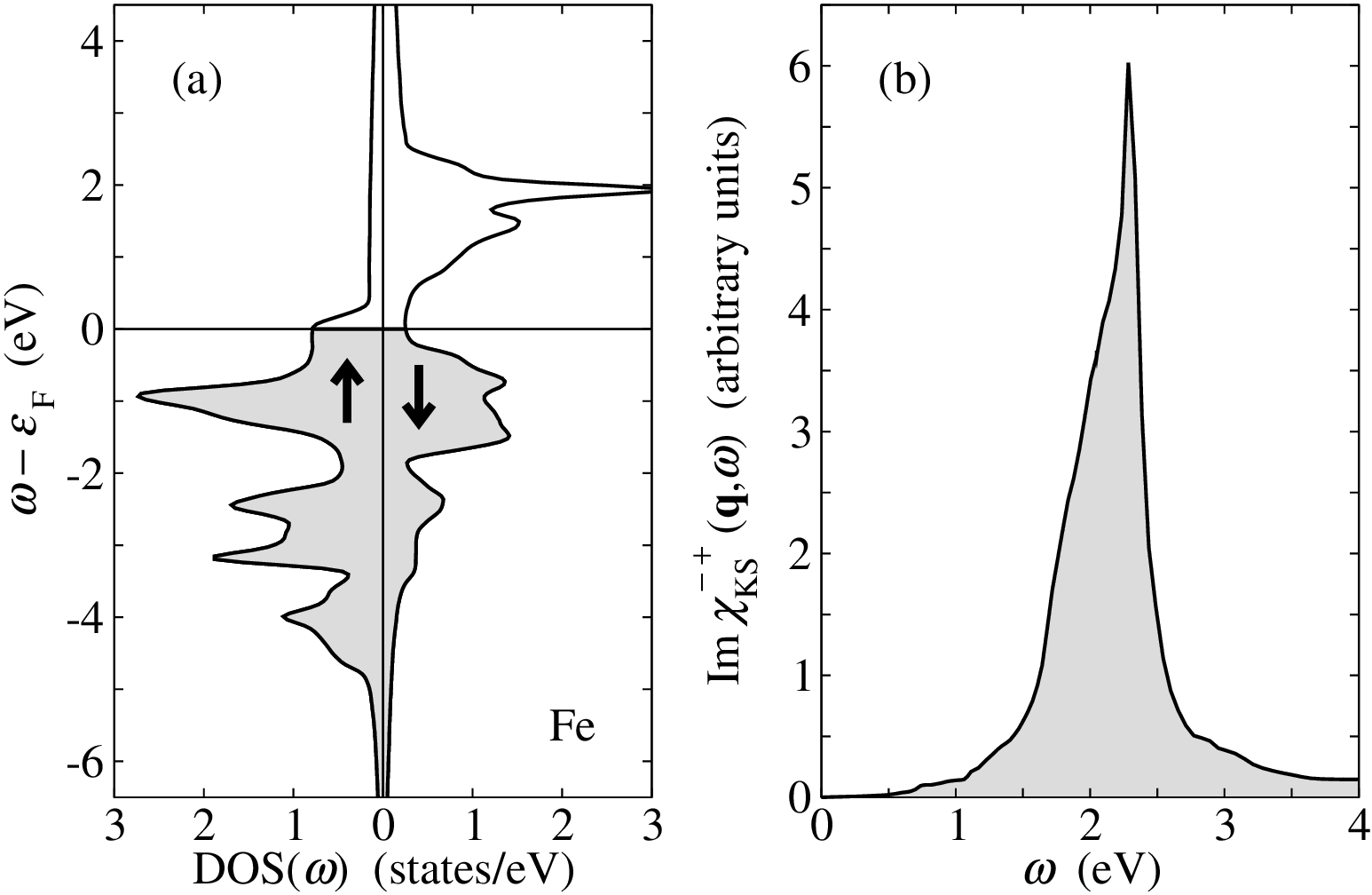}
\caption{(a) Density of states $\mathop{\mathrm{DOS}}(\omega)$ for the majority (up) and minority (down) spin channel of ferromagnetic Fe. (b) Imaginary part of the non-renormalized Kohn-Sham spin susceptibility for $\mathbf{q} = (0.1,0.1,0) 2 \pi / a$.}
\label{Fig:Fe_110_dos_Stoner}
\end{figure}

\begin{figure}[b!]
\centering \includegraphics*[scale=0.35]{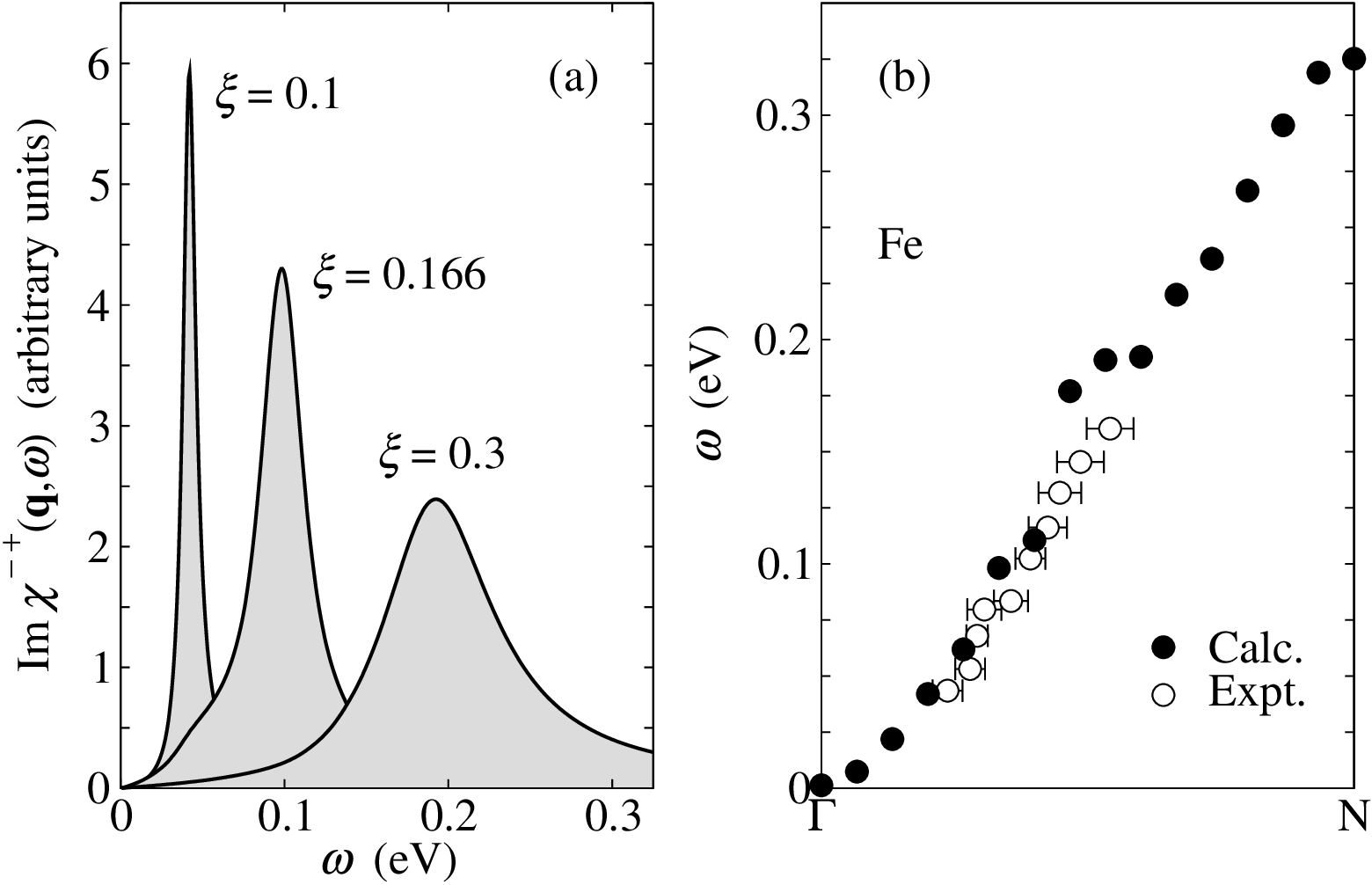}
\caption{(a) Imaginary part of the renormalized spin susceptibility of Fe for wave vectors $\mathbf{q} = (\xi,\xi,0) 2 \pi / a$ and (b) calculated spin-wave dispersion along the [110] direction compared to experimental data from Ref.\ \cite{Loong1984}.}
\label{Fig:Fe_110_magnon_dispersion}
\end{figure}

As an example we show results for iron, obtained with 30$\times$30$\times$30 mesh points in the full Brillouin zone, $G_\mathrm{max} = 4.5\,\mathrm{Bohr}^{-1}$, $L_\mathrm{max} = 4$ and 100 unoccupied bands. Figure \ref{Fig:Fe_110_dos_Stoner} shows the transverse spin susceptibility of the non-interacting Kohn-Sham system. As there is no dynamic correlation in this case, only single-particle Stoner excitations exist. As a consequence, the spectral function $\mathop{\mathrm{Im}} \chi_\mathrm{KS}^{-+}(\mathbf{q},\omega)$ exhibits a peak at around 2\,eV, which equals the exchange splitting visible in the density of states and corresponds to spin-flip transitions between occupied majority and unoccupied minority states. When dynamic correlation is included in the form of the screened Coulomb interaction, an additional spin-wave peak appears at low energies as illustrated in Fig.\ \ref{Fig:Fe_110_magnon_dispersion}(a). Plotting the peak positions as a function of the wave vector yields the spin-wave energies as displayed in Fig.\ \ref{Fig:Fe_110_magnon_dispersion}(b) for the [110] direction in iron. The dispersion obtained in this way, with no empirical parameters, is in excellent agreement with experimental data \cite{Loong1984}.

\section{Summary and outlook}

We have discussed a new implementation of many-body perturbation theory within the FLAPW method. A number of features, such as the copious use of symmetries, the treatment of the Coulomb singularity or expedient basis transformations that allow efficient truncations, which are described in detail elsewhere \cite{Friedrich2009,Friedrich2010}, ensure a high computational efficiency. Our results for electronic excitations and associated spectroscopies in solids are in excellent quantitative agreement with experiments and show that full-potential calculations are now feasible even without plasmon-pole models or other far-reaching simplifications. As the FLAPW method is applicable to transition metals and other complex materials that cannot be easily treated with pseudopotentials, this opens up new prospects for theoretical investigations. We have already started out on this path by exploring spin-wave excitations in ferromagnets. Without any empirical parameters, we obtained spin-wave dispersions in very good agreement with experiments. The rich physics of spin-dependent phenomena means that many further developments are still necessary, however. For example, the relativistic spin-orbit coupling and non-collinear magnetism require appropriate extensions, as do spin dynamics at finite temperature or the inclusion of spin-dependent scattering in the electronic self-energy. Outside the field of nanomagnetism, the linear and non-linear optical properties of modern optoelectronic materials provide another timely challenge.

\section*{Acknowledgement}

We benefitted from useful discussions with G.\ Bihlmayer, M.\ Niesert, A.\ Gierlich, F.\ Freimuth, T.\ Kotani and T.\ Miyake. Financial support from the Deutsche Forschungsgemeinschaft through the Priority Programme 1145 and from the EU's Sixth Framework Program through the Nanoquanta Network of Excellence (NMP4-CT-2004-500198) is gratefully acknowledged.

\end{document}